\documentclass[11pt]{article}

\usepackage{epsfig}
\usepackage{graphics}

\usepackage{amsmath}

\newcommand{\N}{N\raise.7ex\hbox{\underline{$\circ $}}$\;$}

\begin{document}

\title{\bf Pioneer Anomaly and Accelerating Universe \\
as Effects  of the  Minkowski Space Conformal Symmetry \footnote{
To appear in the proceedings of the 5-th International Conference
Boyai-Gauss-Lobachevsky: Methods of Non-Euclidean Geometry in
Modern Physics. Minsk, Belarus, October 10-13, 2006 (without
Sec. 5).}}

\author{L.M. Tomilchik \\
 {\small Institute of Physics,  National Academy of Sciences of
Belarus, 220072 Minsk, Belarus }\\
{\small E-mail: l.tomilchik@dragon.bas-net.by} }
\date{}
\maketitle

\begin{abstract}

On the basis of the nonisometric transformations subgroup of the
$SO(4.2)$ group the nonlinear time inhomogeneity one-parameter
conformal transformations are constructed. The connection between
the group parameter and the Hubble constant is established.
It is shown that the existence of an anomalous blue-shifted
frequency drift is
the pure kinematic manifestation of the time inhomogeneity induced
by the  Universe expansion. This conclusion is confirmed via a
generalization of the standard  Special Relativity clock
synchronization procedure to the space expanding case. The
obtained formulae are in accordance with the observable Pioneer Anomaly effect.
The anomalous blue-shifted drift is universal, does not depend on
the presence of graviting centers and can be, in principle,
observed on any frequencies under suitable experimental conditions.
The explicit analytic expression of speed for the recession
--- intergalactic distance ratio is received in the form of a function
of the red shift $z$, valid in the whole range of its variation.
 In the small $z$ limit this
expression exactly reproduces the Hubble law. The maximum value
of this function at $z=0.475$ quantitatively corresponds to the
experimentally found value $z_{exp}=0.46\pm0.13$ of the
transition from the decelerated to the accelerated expansion of the
Universe.
\end{abstract}

\begin{center}
{\bf 1. Introduction}
\end{center}


 A nature of the blue  frequency shift with a numerical value
 $\frac{d\nu}{dt}=(5,99\pm 0,01)10^{-9}$ Hz/s detected in the signals retransmitted by
 Pioneer 10 and 11 satellites (Pioneer Anomaly - PA) and interpreted as the Doppler shift,
 that is due to the uniform anomalous acceleration towards the Sun amounting to $a_p=(8,74\pm 1,33)10^{-8}$
 cm/s$^2$, still remains not clear (see [1-3]).

The fact that the observable $a_p$ is close to the quantity
$W_0=cH_0$ ($c$ -- speed of light, $H_0$ -- Hubble constant)
points to the cosmological origin of the PA effect. The standard
GR, however, specifies the contribution of cosmological expansion
as a value on the order of $H_0^2$ (rather than $\sim H_0$) [4].
Besides, a sign of this effect is opposite to the observed one
[3].

Nevertheless, an interpretation of PA as a local manifestation of
the cosmological expansion seems to be acceptable when using the
possibilities inherent in the general symmetry group of the
Minkowski space.

Most recent experimental data of astrophysics clearly indicate an
infinitesimal space curvature of the observable Universe. Because
of this, it seems natural to select the Minkowski space as a model
for space-time manifold of the modern Metagalaxy. As is well
known, its most wide symmetry group is the group $SO(4.2)$ that,
apart from  the Poincar\'{e} group isometric transformations,
includes a subgroup of special conformal transformations and
dilatations changing the space and time scales (e.g., see [5]).
Naturally, it may be expected that an expansion of the space-time
symmetry by the inclusion of this subgroup transformations will
result in some generalization of the standard kinematics of the
Special Relativity (SR).

In the present paper on the basis of the nonisometric
transformations subgroup of the $SO(4.2)$ group the nonlinear time
inhomogeneity one-parameter conformal transformations are
constructed (Sec. 2). The connection between the group parameter and the
Hubble constant $H_0$ is established. It is shown that the
existence of an anomalous blue-shifted frequency drift equal to
$\nu H_0$ Hz/s in the location-type experiments performed under
the condition $\Delta t H_0\ll  1$ is the pure kinematic
manifestation of the time inhomogeneity induced by the  Universe
expansion.  The obtained formulae reproduce the
observable Pioneer Anomaly effect. The anomalous blue-shifted
drift is universal, does not depend on the presence of graviting
centers and can be observed on any frequencies under suitable
experimental conditions (Sec. 3). This conclusion is confirmed via a generalization of
the standard  Special Relativity clock synchronisation procedure
to the space expending case (Sec. 4). The explicit analytic expression of speed of recession
--- intergalactic distance ratio is received in the form of the function
of red shift $z$, valid in the whole range of $z$ variation.
 In the small $z$ limit the
expression obtained exactly reproduces the Hubble law. The maximum
in this function at $z=0.475$ quantitatively corresponds to the
experimentally founded value ($z_{exp}=0.46\pm0.13$) of the
transition of the decelerated to the accelerated expansion of the
Universe (Sec. 5).


\begin{center}
 {\bf 2. Conformal transformations and time inhomogeneity}
\end{center}
\hspace{4mm} It is known that the group $SO(4.2)$ considered as a
symmetry group of the Minkowski space, in addition to  the
ten-parameter Poincar\'{e} group, includes also a five-parameter
subgroup of the transformations changing the space-time scales.
This group involves the four-parameter Abelian subgroup $S$ of the
special conformal transformations (SCT). \footnote{Hereinafter the
metric used is $\eta_{\mu\nu}=\text{diag}(1,-1,-1,-1)$.}
$$
x'^\mu = \sigma^{-1}(a,x)\left\{ x^\mu +a^\mu(x^\alpha
x_\alpha)\right\}, \eqno(1)$$ where
$$ \sigma(a,x)=1+2(a^\alpha x_\alpha)+(a^\alpha a_\alpha) (x^\beta x_\beta),
\eqno(2)$$ $ a^\mu$ -- four-vector parameter with a dimension of
the reciprocal length varying in the region $-\infty < a^\mu
<\infty$, and also the one-parameter group of dilatations $D$
$$ x'^\mu = \lambda x^\mu,
\eqno(3)$$ where the positive dimensionless parameter $\lambda$
belongs to the open semi-infinite interval $0<\lambda <\infty$.

$D$ is an external automorphism group of the group $S$.
Consequently, the transformations
$$
x'^\mu = \sigma^{-1}(\lambda a,x)\cdot\lambda\left\{ x^\mu
+\lambda a^\mu(x^\alpha x_\alpha)\right\} \eqno(4)$$ form the
non-Abelian group $G_{SD}$ isomorphic to the semidirect product of
the groups $S$ and $D$ $(G_{SD}\approx S \times )D)$ with a
well-known composition rule for the parameters
$$
(a^\mu_2,\lambda_2)(a^\mu_1,\lambda_1)=(a^\mu_2
+\lambda_2(a^\mu_1), \lambda_2\lambda_1), \eqno(5)$$ \noindent
where $\lambda_2a^\mu_1$ -- image of the element $g(a^\mu_1)$ in
case of automorphism $\alpha(\lambda_2)$, and
$\lambda_2(a^\mu_1)=\lambda_2^{-1}(a^\mu_1)$.

The transformations of (4) do not affect invariance of the light
cone equation (zero interval).

Subsequently, keeping in mind a well-known kinematic SCT
interpretation relating the parameter $a^\mu$ to the uniform
four-acceleration [6], the vector parameter $a^\mu$ is assumed to
be space-like (i.e. it is set to $(a^\alpha a_\alpha)<0)$. For
now, no restrictions are imposed on the vector $x^\mu$.

For simplicity, we select a frame of reference (FR), where the vector
$x^\mu$ is of the form $x^\mu =\left\{ x^0= ct,x,0,0\right\}$,
whereas the four-vector $a^\mu$, due to its space similitude,  may
be of the form $a^\mu =\left\{0,a,0,0\right\}$. In this case the
transformations of (4) lead to a two-parameter group that is
isomorphic to the group of shifts and dilatations in the
two-dimensional subspace $\{ x_0,x\}$ of the Minkowski space, with
a binary operation meeting the composition rule (5):
$$
g(a_2,\lambda_2)\circ q(a_1,\lambda_1)=g(a_2+\lambda_2^{-1}
a_1,\lambda_2\lambda_1). \eqno(6)$$

 As is known, such a group
includes the Abelian invariant subgroup isomorphic to the simple
shear group. The elements of this group are of the form $g(b,1)$,
where $-\infty <b<\infty$. The parameter $b$ as a function of the
parameters $a$ and $\lambda$  may be selected as follows:
$$b=\mp a\eta ,\eqno(7)$$
where
\begin{equation}
\left\{
\begin{array}{l}
1-\lambda^{-1} \;\; if\;\;\lambda >1, \\
1-\lambda \;\; if\;\;\lambda < 1,
\end{array}%
\right.  \nonumber
\end{equation}
$$0\leq\eta\leq 1.$$ As this takes place, the lower (upper) sign
corresponds to expansion (compression) $\lambda >1$ ($\lambda
<1$).

As follows from (4), transformations of the one-parameter group
under study are in the explicit form
$$
x'^0=\sigma^{-1}(b,x)x^0, \quad x'=\sigma^{-1}(b,x)\{
x+b(x^2_0-x^2)\}, \eqno(8)$$ where
$$
\sigma(b,x)=1-2bx-b^2(x^2_0-x^2).$$ Most interesting in this
aspect is a case of the zero-valued interval (isotropic vector
$x^\mu)$ when the relations of (8) take the following form:
$$
x'^{0}=\frac{x^{0}}{1-2bx},\qquad x'=\frac{x}{1-2bx} \eqno(9)
$$ on condition $x=\pm ct$, where the signs ($\pm$)
are associated with a signal propagation in counter directions.
Let us consider the expansion case ($b=-ax$). With due regard for
(7), from the relations of (9) we obtain
$$
t'=\frac{t}{1\pm\frac{t}{t_{max}}}, \eqno(10)$$ where the
parameter $ t_{max}=(2a\eta c)^{-1} $ has the dimension of time,
the signs $\pm$ being associated with a signal propagation in the
forward and backward directions.

As seen, the parameter $t_{max}$ at each fixed value of the
parameters $a,\eta$ sets the upper limit on possible values of
$t'(t)$ when a signal is propagating in the forward (backward)
direction. If it turns out that possible values of the parameter
$|b|=a\eta$ are bounded below by $b_{0}=a_{0}\eta_{0}$, the
permissible time interval is bounded above by
$t_{max}=(2a_{0}\eta_{0}c)^{-1}$. A nonlinear character of the
transformations (10) suggests nonuniformity of time.
Geometrically, these transformations represent a deformation
(respectively compression or extension) of  the light cone
generating lines. And identifying $(t,x)$ with the coordinates of
an event in the ordinary IFR, we find that $(t',x')$ will be
representing the coordinates of the same event in the  deformed
FR. As will be shown later, it may be interpreted as a
special-form noninertial FR. Clearly the transformations of (10)
cause a change in the length of any finite segment of the
light-cone generating lines. We write the required transformations
for the following sequence of events: emission of a signal at the
world point $A$ at the instant of time
 $t'_{A}$; its tranfer to the world point $B$, instantaneous reemission, and subsequent return to the world point $A$ at the instant of time $t_{A}$. That is to say, we have
$$
t'_B-t'^0_A=\frac{t_B-t^0_A}{1+\frac{t_B-t^0_A}{t_{max}}}, \qquad
\eqno(11)
$$
$$
t'_A-t'_B=\frac{t_A-t_B}{1-\frac{t_A-t_B}{t_{max}}}. \eqno(12)
$$
Here the primes denote the coordinates of the associated events in
the deformed FR. As seen, for the coincident time intervals
$t_A-t_B$ and $t_B-t^0_A$ the time intervals $t'_A-t'_B$ and
$t'_B-t'^0_A$ are not equal to each other, and  $t'_A-t'_B >
t'_B-t^0_A$. This disagreement should result in the experimentally
observed results; primarily in the experiments on location of the
space-separated objects by electromagnetic signals. In experiments
of this type the measured time of the whole process is determined
as a time interval between the transmission of a signal ($t^0_A$)
and its return ($t_A$) because $\Delta t_{obs}=t_A-t^0_A$. Since
time coincidence is assumed both in the forward and backward
direction, the time interval required for a signal to cover a
distance to the object equals
$$
\Delta t=\frac{1}{2}(t_A-t^0_A), \eqno(13).
$$
\noindent And the distance itself, due to the universal constancy
of a signal rate (first of all its independence of the source
motion rate), is determined by the following relation:
$$
\Delta r=c \Delta t. \eqno (14).
$$

\begin{center}
{\bf 3.Time inhomogeneity and Pioneer Anomaly}
\end{center}

For the comparison with experiment, we consider a small-time
approximation when the condition $t/t_{max}\ll 1$ is valid. In
this case, restricting ourselves to consideration of the
first-order infinitesimal terms, from (12) we obtain
$$
\Delta t'=\Delta t  , \eqno(15)$$
$$
t'_B-t'^0_A=\Delta t -\frac{(\Delta t)^2}{t_{max}}, \qquad
t'_A-t'_B=\Delta t +\frac{(\Delta t)^2}{t_{max}} . \eqno(16)$$

As follows from formulae (15) and (16), the time interval $\Delta
t'$ determined by the relation of (15) is smaller than the actual
length of time $t'_A-t'_B$ required for a signal to cover the
distance from the source to the observation point. It is easily
seen that this results in the observable radiation-frequency
shift, in the direction of increasing values, in accordance with
the following relation:
$$
\nu_{obs}=\nu (1+\frac{2\Delta t}{t_{max}}). \eqno(17)$$ \noindent
Here $\nu$-- frequency of the signal from a stationary, in the
ordinary sense, source, $\nu_{obs}$ -- really observed frequency.

Whence the expression for a relative frequency shift in unit time
is as follows:
$$
\frac{\nu_{obs}-\nu}{\Delta t}=2t^{-1}_{max}\nu=4a_0\eta_0c\nu.
\eqno(18)
$$

Within the scope of the approach under study, the blue  frequency
shift is universal, and in conditions of the expanding space-time
manifold it should be observed, in principal, at all frequencies.
Proceeding from this point of view, the Pioneer Anomaly is the
first experimentally recorded effect of this type that may form a
basis for experimental estimation of the numerical value of
$t_{max}$.

It is known (see [1-3]) that the directly observable effect of PA
consists in revealing of the uniform rate of a positive shift
$\dot{\nu}_{obs}$ at the operating frequency $\nu =2,29\cdot 10^9$
Hz for satellite repeaters by the amount calculated as $$
\dot{\nu}_{obs}=(5,99\pm 0,01)\cdot 10^{-9} \text{Hz/s}.
\eqno(19)$$

Selecting for a minimum value of the parameter $a$ the quantity
$H_0/c$ ($H_0=2,4\cdot 10^{-18} \text{s}^{-1}$ -- Hubble
constant), assuming in (18) $\nu =2,29\cdot 10^9$ Hz and
$\frac{d\nu'}{dt}=\dot{\nu}_{obs}$, by comparison with the
numerical value $\dot{\nu}_{obs}$ from (19), we can find that for
$\eta_0\cong 0,272$ equation (17) reproduces the observable PA
effect, both quantitatively and qualitatively. And as this takes
place, the value of the parameter $t_{max}$ is equal to
$t_{max}=(2\eta_0 H_0)^{-1}\cong 1,84\cdot H^{-1}_0$.

It should be noted that, within the scope of the proposed
approach, the effect is purely kinematic (in the same sense as the
effects of time "retardation" and Lorentzian compression are
kinematic in a standard Special Relativity). Because of this, the
approach requires no specific dynamic substantiation. In fact, the
question is about the exhibited time inhomogeneity induced by the
cosmological expansion and described in the general case by
equation (10). In the approximation of $H_0t\ll 1$, valid in a
very wide time interval, this inhomogeneity is exhibited as a term
quadratic in $t$, that appears in the expression determining the
nonlinear dependence of $t'$ on $t$.

For the distances $\Delta r'=ct'$ covered by a signal along the
light-cone generating lines in the forward and backward directions
we obtain the following expression:

$$
\Delta r'_\pm =c\Delta t\mp \frac{c(\Delta t)^2}{t_{max}}= c\Delta
t \mp \frac{W_0(\Delta t)^2}{2}. \eqno(20)$$ As is demonstrated,
in conditions of the expanding space-time manifold (in the
small-time approximation $t\ll H_0^{-1}$) the distances covered by
a signal along the light-cone generating lines in the forward and
backward directions are distinguished from $r_0=c\Delta t$ by the
quantity $\delta r=\frac{W_0t^2}{2}$, where
$$
W_0 = 2ct^{-1}_{max}=4a_0\eta_0c^2 \eqno(21)$$ is a constant with
the dimension of acceleration.

From the conventional point of view, the situation seems to be
caused by a radiation source experiencing at all space points the
uniform acceleration $W_0$ directed  to the origin of coordinates
(observation point).

Substituting into (21) the numerical values $c=3\cdot 10^{10}$ cm,
$H_0= 2.4\cdot 10^{-18}$ s$^{-1}$, $\eta_0=0.272$ for the
acceleration $W_0$ we obtain the numerical value $W_0\cong
7.83\cdot 10^{-8} \text{cm/s}^2$ that is in a good agreement with
the experimentally recorded value $a_p=(8,74\pm 1.33)\cdot 10^{-8}
\text{cm/s}^2$ cited in [1-3].

Note that the uniform acceleration $W_0$ is actually background in
character. Consequently, one may consider that it is
characteristic for "expanding" Frame of Reference (FR) regarded as
a specific case of the noninertial one. As this takes place, the
corresponding four-acceleration $W_\mu$ is given by  the
four-vector having in the co-moving FR the following form: $W_\mu=
\left\{ 0,W_0\frac{\mbox{\boldmath $r$}}{r}\right\}$. And the
параметр $a^\mu$ is related to $W^\mu$ by the relation
$a^\mu=\frac{1}{c^2}W^\mu$ that is adopted in the concept
associated with a well-known kinematic interpretation of SCT (see
[6]).

According the proposed approach the appearance of the quadratic in
time term in the formulae (16) and (20) under condition
$t/t_{max}<<1$ mimics the effect of the constant acceleration (21)
directed in the accordance with (20) towards the point of
observation. It is easy to see that the condition $t/t_{max}\sim t
H_0<<1$ is sure satisfied in any real location-type experiment in
our solar system.

\begin{center}
{\bf 4. The clock synchronization procedure in expanding space}
\end{center}

Let us demonstrate that the same result may be acquired when we
consider a standard location procedure for the clock
synchronization in conditions of the expanding space-time
manifold.

It is common knowledge that the kinematic physical basis for the
SR concept is formed using the location procedure for
synchronization of the space-separated clocks at rest in each
fixed IFR (see [7]).
 In so doing it sands to reason that the condition used should set an invariable distance between
 each clock pair in the process of synchronization.  As shown in [8],
 with a change in the space scales the synchronization condition is
 modified due to inclusion of the cosmological expansion.

When the space scales are varying in accordance with the
postulates of a homogeneous and isotropic model, it can be assumed
that the radial distances $R$ have the exponential time-
dependence
$$
R(t)=R(t_0) \exp \left\{ \pm \int^t_{t_0} H(\tau)d\tau \right\} .
\eqno(22) $$
 Here $H$ -- Hubble parameter that is generally
dependent on time, and the signs of $(\pm)$ are associated with
expansion and compression. Hereinafter our consideration is
confined to the expansion case only.

Let the identical clocks located at the points $A$ and $B$ be
separated by the space interval $R _{AB}$. The synchronization
procedure remains the same as previously: a synchronizing signal
is transmitted from the point $A$ at the instant of time $t^0_A$,
reflected at the point $B$ at the instant of time $t_B$, and
returned back to the point $A$ at time $t_A$. Now the distances
$R_{AB}$ and $R_{BA}$ covered by the signal in both directions are
different ($R_{BA} > R_{AB}$).

If $c_1$ and $c_2$ -- average rates of the synchronizing signal in
the "forward" and "backward" \hspace{5mm}
directions, with the use
of the postulate that the synchronizing-signal rate is independent
of the source motion rate, the following relations may be written:
$$
R_{AB}= c_1 (t_B -t^0_A),\quad R_{BA}= c_2 (t_A -t_B). \eqno(23)$$
Taking them into account, we have
$$
\frac{R_{BA}}{R_{AB}} = \frac{c_2(t_A-t_B)}{c_1(t_B-t_A^{0})}=\exp \left\{
\bar{H}(t_A-t_B)\right\}, \eqno(24)$$ where
$\bar{H}=(t_A-t_B)^{-1}\int^{t_B}_{t_A}{H dt}$.

As the right-hand side of (24) is above 1, the Einstein conditions
\mbox{$t_A -t_B =t_B-t^0_A$} and $c_1=c_2$ may be simultaneously
fulfilled only in the absence of expansion. And in the case under
study there is an alternative:
$$ \text{(I)}~ t_A -t_B > t_B -t^0_A, \quad c_1 = c_2;\quad
 \text{(II)}~ t_A -t_B = t_B -t^0_A, \quad c_2 > c_1.  $$
With due regard for a key role played in the SR concept and modern
metrology by the representation for the universal stability of the
synchronizing signal rate (i.e. constant $c=3\cdot 10^{10}$ m/s),
we choose the first variant (as variant II has been considered in
[8]).

On condition $c_1=c_2=c$, from the relation of (24) we derive the
following expression:
$$
\frac{t_A-t_B}{t_B-t_A^0}=\exp\left\{ \bar{H}(t_A-t_B)\right\},
\eqno(25)$$ that comprises an transcendental equation for the
definition of the unknown $t_B$ in terms of the given values $t^0_A$
and $t_A$. Under the condition $\bar{H}(t_A-t_B)<<1$  (25) leads to
the relation
$$t_A-t_B=(t_B-t^0_A)\{ 1+\bar{H}(t_A-t_B)\}
\eqno(26)$$ representing the following quadratic equation for
$t_B$:
$$
t_B^2-\frac{2}{\bar{H}}\left\{
1+\frac{\bar{H}}{2}(t_A+t^0_A)\right\}t_B
+\frac{1}{\bar{H}}(t_A+t^0_A)+t_At^0_A=0.$$ Its solutions, in the
approximation linear with respect to $\bar{H}(t_A-t^0_A)$, are of
the form
$$
t_B(\pm)= \frac{1}{\bar{H}}+\frac{t^0_A+t_A}{2}\pm
\frac{1}{\bar{H}}\left\{
1+\frac{\bar{H}^2(t_A-t^0_A)^2}{8}\right\}. \eqno(27)$$ Proceeding
from the correspondence principle, in accordance with which for
$\bar{H}=0$ the result of the standard SR should be reproduced by
$t_B=\frac{1}{2}(t^0_A+t_A)$, we have to choose the solution
$t_B(-)$. Then we get
$$
t_B=\frac{1}{2}(t^0_A+t_A)- \frac{\bar{H}(t_A-t_A^0)^2}{8}.
\eqno(28)$$ As seen, $t_B$ has a linear dependence on the
difference in indications $(t_A-t_A^0)$ of the "reference"
\hspace{2mm} clocks and, within the approximation used, includes
an additional quadratic term.

For the propagation time of a synchronizing signal in the forward
$(t_{AB}=t_B-t_A^0)$ and backward $(t_{BA}=t_A-t_B)$ directions we
find that

$$
t_{AB}= t-\frac{\bar{H}t^2}{2}, \quad
t_{BA}=t+\frac{\bar{H}t^2}{2}, \eqno(29)$$ where
$t=\frac{1}{2}(t_A-t_A^0)$ is the propagation time of a
synchronizing signal as used in the standard SR.

From the relation of (29) we derive the following expressions to
determine the distances covered by a signal in the forward and
backward directions:
$$
R_{AB}=r_0-\Delta r=ct-\frac{c\bar{H}t^2}{2},\quad
R_{BA}=r_0+\Delta r=ct+\frac{c\bar{H}t^2}{2}. \eqno(30)$$ From the
customary point of view, the situation looks as though in the
process of synchroniza\-tion the clock $B$ were moving with
respect to the "reference" \hspace{2mm}  clock $A$ with the
acceleration, equal to
$$
W=c\bar{H} \eqno(31)$$  and directed towards the clock $A$ (i.e.
to the observation point).

 It should be noted that the presence of a term quadratic with respect to $t$ in the relation
(29) makes it possible to suggest the "clock acceleration"
\hspace{5mm} effect as $\frac{d^2t_{AB}}{dt^2}=\bar{H}\neq 0$
(see[9] and reference [9] ibidem).

In essence, the expressions (29) and (30) thus obtained are
coincident with the relations (16) and (20) providing an identity
of $\bar{H}$ with the Hubble constant $H_0$ and assuming
$\eta_0=\frac{1}{4}$.

It is clear that a choice of the sign (-) in the initial relation
(22) - compression case - leads to the reciprocal substitution of
signs in the resultant relations (29) and (30), that is an
additional argument in support of the sign choice  in definition
(7) of the parameter $b$. The frequency shift in case of the
compressed space-time manifold would turn to be red, and the
associated "minimum acceleration" were directed from the
observation point.

It is interesting that the real measuring procedure performed in
the process of tracking the satellite motion completely replicates
the principal features of clock synchronization in SR: signal
transmission, its repeating by a satellite transmitter, and return
to the observation point.

\begin{center}
{\bf 5. The  dependence of distance on red shift. The Hubble law}
\end{center}

Let us consider the experimentally verifiable cosmological consequences of the
conformal time transformations (10) which are not restricted to
the case of small values of $t/t_{max}$.
First of all we  show that the explicit expression for the distance between
the signal emitter and the point of observation can be derived
directly from conformal time transformations (10) in the form of a
simple function of the red shift $z$. For this propose we consider
the case of a signal traveling in the direction of negative
generatrix of light cone (lower sign in the formula (10)):
$$
t'=\frac{t}{1-\frac{t}{t_{max}}}. \eqno(32)
$$
For small time increments $\Delta t'$ and $\Delta t$  we have from
(32)
$$
\Delta t'=\Delta t(1-\frac{t}{t_{max}})^{-2}. \eqno(33)
$$
If $\Delta t$ and $\Delta t'$ are the periods of oscillations of
emitted ($\Delta t= T_{emitted}$) and received ($\Delta
t'=T_{observable}$) signals correspondingly, then using the
standard definition of red shift
$$
\frac{\lambda_{observable}}{\lambda_{emitted}}=z+1, \eqno{(34)}
$$
where  $\lambda_{observable}=cT_{observable}$ and
$\lambda_{emitted}=cT_{emitted}$,  we find from (33) the
expression
$$
\frac{\lambda_{observable}}{\lambda_{emitted}}=
(1-\frac{t}{t_{max}})^{-2}=z+1, \eqno(35)
$$
which gives
$$t(z)=t_{max}\frac{(z+1)^{1/2}-1}{(z+1)^{1/2}}.\eqno(36)$$
Here $t(z)$ represents by definition the time interval between the
moments of emitting and receiving the light (electromagnetic)
signal. So, supposing that the speed of light is constant and does
not depend on the velocity of the emitter, the quantity $R=ct$ can
be regarded as the distance covered by the signal.

From the formula (36) we obtain an expression which determines an
explicit form of dependence of $R$ on the red shift
$z$\footnote{Relation of the type $R(z)=const\cdot z$, that connect
cosmological distance $R$ with the red shift, seems to be firstly
obtained from the conformal symmetry arguments  by Ingraham [10] in
1954.}:
$$
R(z)=R_u
\frac{(z+1)^{1/2}-1}{(z+1)^{1/2}}=R_u\left(1-\frac{1}{(z+1)^{1/2}}\right).
\eqno(37)
$$
Here$R_u=ct_{max}$ is a parameter, which, within in the model
suggested, has the sense of the limit (maximal) distance. We will
assume its value to be equal to
$$
R_u=2cH^{-1}_0. \eqno(38)
$$

Quantity $R(z)$ defined by (37) corresponds to the distance, which
in   cosmology is referred to as a location distance. In principle
the  relation (37) allows direct experimental verification in the
whole range of $z$ variation and can be confirmed or refused by
observations.

To obtain explicit expression for the Hubble law, we make use of the
well-known formulae describing the Doppler effect in Special
Relativity. However in the context of our approach, to find the
explicit expression for the longitudinal Doppler effect, it is
convenient to apply the formulae which immediately follow from the
Lorentz boosts written in the light-cone variables:
$$
u=\frac{1}{2}(x_0+x),\;\;\;\; v=\frac{1}{2}(x_0-x).
$$
These expressions are (see, for example, [11]):
$$
u'=k(\beta)u,\qquad v={k}^{-1}(\beta)v, \eqno(39)
$$
where $k(\beta)=(\frac{1-\beta}{1+\beta})$,  $\beta=V/c$, $V$ is
the velocity which can be identified as the radial component of
relative velocity of emitter and receiver motion.

Clearly, in terms of $u$ and $v$, the Lorentz boosts have the form
of dilatations. For the description of the Doppler effect  we need
use the equation of light cone $4uv=0$. Then for the case of the
signal traveling in positive ($v=0$) and negative ($u=0$) directions
we have in coordinates ($x, t$):
$$
t'=(\frac{1\mp\beta}{1\pm\beta})^{1/2} t, \eqno(40)
$$

In the case of emitter and receiver moving away from each other,
we find from (40) for small time increments $\Delta t'$ and
$\Delta t$:
$$
\Delta t'=\Delta t \left(\frac{1+\beta}{1-\beta}\right)^{1/2},
$$
whence, taking into account (40), the known expression for
function $V(z)$ follows:
$$
\frac{V(z)}{c}=\frac{(z+1)^2-1}{(z+1)^2+1}. \eqno(41)
$$
Equation (A10) is valid in the whole range of velocity $V$
variation.

In the approximation $z\ll1$  we obtain from (37) and (40) to
accuracy of the  terms $\sim z^2$)  $V(z)/c\cong z$ and
$R(z)/R_u\cong z/2$, whence, taking into account (38), the
expression for the Hubble law in its standard form follows:
$$V=H_0R.\eqno(42)$$
In the general case  we find by (37) and (41) the following
expression for the ratio $V/R$:
$$
\frac{V(z)}{H_0
R(z)}=f(z)=\frac{1}{2}\frac{(z+1)^{1/2}}{(z+1)^2+1}
\cdot\frac{(z+1)^2-1 }{(z+1)^{1/2}-1}. \eqno (43)
$$
Clearly, $\lim_{z\rightarrow 0} f(z)=1$ and $\lim_{z\rightarrow
\infty} f(z)=1/2$. Asymptotic behavior of the function $f(z)$ is
determined as follows:
$$
f(z)|_{z\ll1}=1+z,\qquad
f(z)|_{z\gg1}=\frac{1}{2}(1+\frac{1}{\sqrt{z}}).
$$

\begin{figure}[h!]
\centering
\epsfig{file=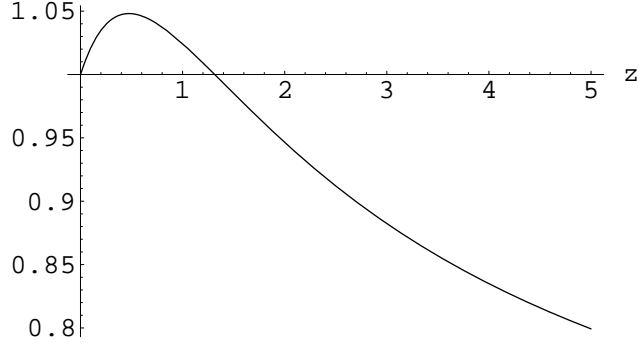,width=8.5cm}
\caption{The function $f(z)$. }
\end{figure}

\begin{figure}[h!]
\centering
\epsfig{file=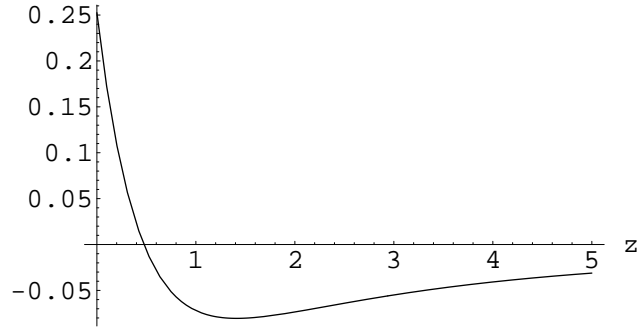,width=8.5cm}
\caption{The derivative $f'(z)$. }
\end{figure}

\begin{figure}[h!]
\centering
\epsfig{file=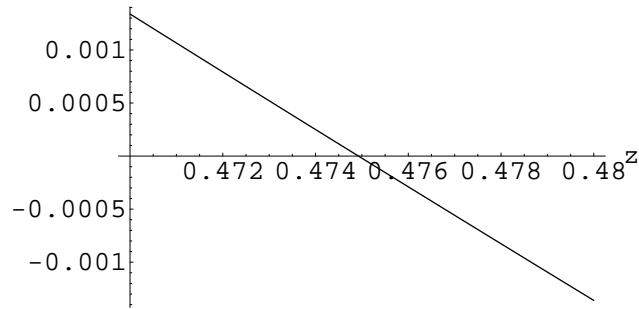,width=8.5cm}
\caption{The position of the zero of  $f'(z)$. }
\end{figure}

The function $f(z)$ and its derivative $f'(z) = \frac{d f(z)}{dz}$ are shown
in Fig. 1 and Fig. 2. Obviously, $f(z)$ has a maximum at $z_0\cong0.475$
(see Fig. 2 and Fig. 3). Overall variation of $f(z)$ demonstrates that
in the interval $0\leq z<z_0$ the distance $R(z)$ increases with $z$
more slowly, and in the interval $z_0<z<\infty$ approaches its limit
value $R_u$ more rapidly, than the velocity $V(z)$ approaches its
limit $c$. Horizontal line in Fig.~1 represents strict Hubble law (42).

As
regards a  possible treatment of the behavior of the function
(43) in terms of the standard approach using the decelaration
parameter, we are to notice  the following. According to the
 pure kinematic approach proposed in our paper,  the source of the
effects induced by the cosmologic expansion is the time conformal
inhomogeneity.  The "acceleration" attributed to the
emitting source arises because of treating the actual nonlinear time
dependence $t'(t)$ in terms of the traditional theoretical paradigm
based on the time homogeneity concept. The uniform "acceleration"
$W_0=cH_0$ which appears in the formula (21) in the $t/t_{max}\ll 1$
approximation is not a "true" but the "effective" acceleration in
the reality. The "acceleration" in the general case must be treated as time-dependent one.
Pure formally it can be determined as  the second derivative of the function
$$
R(t)= c t'= c t\left(1-\frac{t}{t_{max}}\right)^{-1} \eqno{(44)}
$$
and it possess the following form
$$
W(t)=\frac{2c}{t_{max}}\left(1-\frac{t}{t_{max}}\right)^{-3} .
\eqno{(45)}
$$
The "time-dependent effective acceleration" is directed to the
point of observation and coincides in the first approximation in
$t/t_{max}$ with $W_0=cH_0$. This "acceleration" can be presented
according to (35) in the form
$$
W(z)=W_0\left(1+z\right)^{3/2}.
$$
One can say that the numerical value of such an "acceleration"
decreases during the process of the Universe expansion starting from
the very large magnitude. Evidently, the interpretation of $f(z)$
behavior from the position of common treatment  seems as follows. In
the interval $\infty > z > z_0$ there is the \underline{deceleration}
of cosmological expansion, and at the point of $z_0\cong 0{,}475$ it
changes to the \underline{acceleration}. Numerical value of  $z_0$
agrees quite well with experimentally founded "point of change"
$z_{exp}=0.46\pm0.13$ from the deceleration of cosmological
expansion to the acceleration.

At last, we again emphasize the essentially kinematic nature of
the relation (37). It is the manifestation of the
\underline{nonlinear} time deformation (10) which follows from
Special Conformal Transformations exactly in the same manner, as
the Doppler effect follows from the \underline{linear} time
deformation (40), which arise from the Lorentz boosts leaving the
equation of light cone unaltered.

It should be emphasized that the basic formula (10) for the
conformal transformations of the time, as well as all its
consequences are valid on the assumption that the Hubble parameter
$H_0$ is constant. Hopefully this assumption is reasonable as
applied to at least later  stages of the Universe evolution. In
this case the proposed formulae (37) and (43) can be valid for
the experimentally  obtained values of the read shift having the
order of several units.

\begin{center}
{\bf 6. Conclusion}
\end{center}

Provided the proposed interpretation of PA is correct, the
frequency shift measured may be considered as a new independent
high-precision measurement of the Hubble constant.

Moreover, the concept as a whole may be directly tested by the
experiment. The use of the appropriate sources and monochromatic
detectors at required experimental conditions enables observation of
the described anomalous blue shift in its "pure form", when the
source and detector are mutually motionless, practically at all
frequencies.  As according to (18) the effect is linearly growing
with the frequency, it is expedient to use high-frequency radiation.
In principle, this allows for a considerable decrease in the
observation time. To study this effect, it is required to provide
(i) maximum immobility of the source and detector, (ii) elimination
of the gravitational field effect of massive bodies, (iii)
minimization of thermal fluctuations and mechanical deformations.
One should not rule out a possibility of providing all these
conditions in zero gravity at the satellites orbiting along the
circumterrestrial orbits. This proposal has been made in [12].

By the proposed approach, the anomalous blue-shifted drift is a pure
kinematic effect arising from homogeneous and isotropic expansion
\emph{of the flat} four-dimensional Minkowski manifold. In a sense,
this approach is phenomenological with respect to the General
Relativity, in itself being a certain natural (and minimal)
extension of the SR concept, with evident adherence to the
correspondence principle. And most important is the problem
concerning modification of the standard relativistic kinematics
necessitating further investigation.

It should be emphasized that, according to the suggested
interpretation, the effect should be independent of any gravitating
centers present. In other words, the acceleration $W_0\sim cH_0$ is
background in character and can be treated as the manifestation of
the noninertial character of local frame of reference associated
with any point of expanding Universe. It would be noted that such a
"noninertiality" can mimic in accordance with the equivalence principle
the dark energy effect. It is
highly probable that a similar origin may be also attributed to a
minimum acceleration $\sim 10^{-8}$ cm/s$^2$ (Milgrom parameter)
used as a fundamental dimensional constant in the MOND concept (see
[13-15]). Also, note that the parameter $W_0$ arise as the so-called
Mass Dependent Maximal Acceleration (MDMA) (see [16] and references
ibidem), if  we
substitute the value $M=M_u=\frac{4\pi}{3}R^3_u\rho_c$ that is equal
to the Metagalaxy mass ($R_u=cH_0^{-1}$ -- "radius" of the Universe,
$\rho_c=\frac{3H_0^2}{8\pi G}$  - critical density)   into the expression
$W(M)=F_0/M$ for  MDMA, where
$F_0=c^4/G$ ($G$ - gravitational Newton constant) is the Maximum
Force (MF) introduced in [16] and independently in [17,18].  All these
problems call for special consideration in further publications of
the author.

\begin{center}
{\bf  Acknowledgements.}
\end{center}
\hspace {3mm} The author would like to acknowledge E.V. Doktorov, V.V. Kudryashov,
 A.E. Shalyt-Margolin,  Ya.M. Shnir, I.A. Siutsou  and Yu.P. Vybly
   for their invaluable critical
remarks and for the fruitful discussions.

\vspace {3mm}

 The paper is in part supported by the Belarus
Foundation for Fundamental Research (BFFR), Project F06-129.

\end{document}